# Optical disassembly of cellular clusters by tunable "tug-of-war" tweezers

**Anna Bezryadina,[1] Daryl Preece,[2,3] Joseph C. Chen,[4] and Zhigang Chen[1,5]**

[1]Department of Physics and Astronomy, San Francisco State University, San Francisco, CA 94132, USA

[2]School of Mathematics and Physics, The University of Queensland, Brisbane, QLD 4072, Australia

[3] Department of NanoEngineering, University of California, San Diego, La Jolla, CA 92093, USA

[4]Department of Biology, San Francisco State University, San Francisco, CA 94132, USA

[5]The MOE Key Laboratory of Weak-Light Nonlinear Photonics,and TEDA Applied Physical Institute and School of Physics, Nankai University, Tianjin 300457, China

Corresponding author: Z. Chen, Email: *zhigang@sfsu.edu*

Running title: "Tug-of-war" optical tweezers for biofilm test

## Abstract

**Bacterial biofilms underlie many persistent infections, posing major hurdles in antibiotic treatment. Here, we design and demonstrate "tug-of-war" optical tweezers that can facilitate assessment of cell-cell adhesion - a key contributing factor to biofilm development, thanks to the combined actions of optical scattering and gradient forces. With a customized optical landscape distinct from that of conventional tweezers, not only can such tug-of-war tweezers stably trap and stretch a rod-shaped bacterium in the observing plane, but, more importantly, they can also impose a tunable lateral force that pulls apart cellular clusters without any tethering or mechanical movement. As a proof of principle, we examined a *Sinorhizobium meliloti* strain that forms robust biofilms and found that the strength of intercellular adhesion depends on the growth medium. This technique may herald new photonic tools for optical manipulation and biofilm study, as well as other biological applications.**

# 1. Introduction

Antimicrobial resistance is a troubling and growing public health threat[1]. In addition to resistance mechanisms that can develop at the cellular level, the propensity of bacteria to form biofilms further protects them from environmental assaults, including antibiotics and the host immune system[2-4]. In fact, bacterial surface-attachment and subsequent biofilm formation are considered important hallmarks of the capacity of microbial communities to cause persistent infections. Understanding factors that contribute to bacterial aggregation during biofilm maturation is critical to the study of microbial physiology and ecology, as well as to the advancement of new treatments for chronic infections and novel strategies to prevent biofilm-associated problems[5,6]. Thus far, the primary optical tool for studying biofilms has been laser scanning microscopy, from single-photon, two-photon to multiphoton excitation microscopy[7]. While these imaging techniques have reified our current view of complex and heterogeneous biofilm structures, new optical tools are still desired for characterization and manipulation of biofilms, especially since their structures vary over time and under different environmental conditions[8].

Optical tweezers with fine-shaped light beams provide excellent tools for trapping and manipulating bacteria as well as micro- and nano-scale particles[9-20]. Over the past decades, optical tweezers have been routinely used for single-molecule force spectroscopy to understand the mechanics of biological processes. Recently, they have also been employed to study bacterial aggregation, as well as to better characterize bacterial motility and flagellar rotation[21-25]. In conventional single-beam gradient optical tweezers[9] (Fig. 1a), however, a rod-shaped object or bacterium (such as a *Bacillus thuringiensis* cell) tends to align preferentially towards the propagation direction of a trapping beam, preventing direct in-plane observation and manipulation of the trapped object. In single-molecule force measurements, a molecule of interest is often tethered to a surface at one end and attached to a trapped dielectric bead at the other end, or both ends are attached to simultaneously trapped beads in dual-beam ("dumbbell") optical tweezers[13,26,27]. At the single-cell level, a dual-beam optical trap (Fig. 1b) is also necessary for orientation and manipulation of individual cells in space[23]. The need for active control of single or numerous trapped objects has motivated the development of multi-trap optical tweezers, including dynamic holographic tweezers and those created with complex beam shaping techniques[28-32]. Nevertheless, available methods for optical trapping and manipulation of bacteria still have substantial limitations in quantitative characterization of bacterial motility and intercellular interaction. For example, dual-beam optical tweezers rely on a pair of perpendicularly polarized beams and significant user control of each trapping beam: a rod-shaped bacterium has to be trapped and flipped by one of the beams first, and then the user needs to manually control the other trap to orient the cell into the desired observing plane. In addition, the use of optical tweezers to stretch a bacterial cell directly or to break up cellular clusters still remains a challenge.

In this work, we describe an optical tweezer-based assay for the study of bacterial adhesion, relying on a "tug-of-war" (TOW) design from novel shaping of light (Fig. 1c). Instead of using two separate traps under independent control, we "split" a single beam into a pair of elongated trapping beams propagating with a diverging angle. This TOW design has the following advantages over the conventional single- or dual-beam optical tweezers. Firstly, it allows for stable in-plane trapping of a rod-shaped object with a single control implemented at will, without any mechanical movement or phase-sensitive interference. Secondly, and more importantly, such TOW tweezers can apply a tunable lateral pulling force on the trapped object, and the strength of the pulling force can be varied by changing the trapping beam intensity from femto- to pico-newton levels. As an example, we employed the TOW tweezers to trap, stretch, and even break apart *Sinorhizobium meliloti* cellular clusters aggregated under different conditions. We estimated the force needed for disassembly of adhesive *S. meliloti* cells, and found that the strength of bacterial adhesion is dependent on the growth medium. We believe this technique can bring about new avenues of exploration for optical manipulation and biophotonics.

## 2. Materials and Methods

First, we discuss the design and demonstration of the TOW optical tweezers. Conceptually, the design of TOW optical tweezers relies on nontrivial shaping of a Gaussian beam into two elongated (stripe-like) beams with opposite transverse momenta. When applied in the optical tweezers setting, pulling forces arise on both sides of the trapped object, as in a "tug-of-war" duel. In practice, the size, separation, and propagation direction of the two beams can be varied at will, as implemented by encoding the holographic information onto a LabView-assisted spatial light modulator (SLM), thus allowing interactive control of the directions and magnitudes of the pulling forces (see Supplementary Information).

To better visualize the intensity distribution and structure of the resulting TOW beam, a technique for volumetric representations of holographic optical traps is used[33]. By acquiring a stack of two-dimensional images of the trapping beam near the focal plane, experimentally recorded data are replotted in Fig. 1d, in which a composite picture shows the side-view of the beam structure, along with the calculated vector field distribution of the intensity gradient. This design of the TOW tweezers, although still based on the holographic principle, leads to an effective optical tool for trapping rod-shaped objects. Distinct from conventional dual-beam or holographic tweezers, the intensity landscaping in the TOW tweezers manifests a strong intensity gradient in the central "pulling" region, with maximal momenta oriented in opposing directions as a result of the synergistic action of optical scattering and gradient forces. In addition, the two elongated beams in the TOW tweezers provide a better match to the bacterial morphology, enabling stable trapping of a rod-shaped bacterium even at low power levels, thus reducing the effects of photodamage on the trapped cell[9]. The vector field of the intensity gradient of the trapping beam in Fig.1 represents a useful description of the contribution from the gradient force that normally dominates in optical tweezers[34]. By reshaping the trapping beam (thus the force distribution), TOW tweezers can be optimized to trap rod-shaped objects of different sizes and compositions, including silica rods (used as a proof of principle), *Escherichia coli* (Gram-negative bacterium about 2 μm in length), and *B. thuringiensis* (Gram-positive bacterium with cell length ranging from 5 to 15 μm). In fact, TOW tweezers can be reconfigured to trap asymmetrically shaped particles with lengths varying from micrometer to sub-millimeter. As an example, we show in the bottom panels of Fig. 1d experimental snapshots obtained from the TOW trapping and self-aligning of a dividing, rod-shaped *B. thuringiensis* cell. After entering the trap, the bacterium is reoriented onto the observing plane and stretched from two ends, unable to escape the trap.

3. Results and Discussion

To illustrate the feasibility and potential application of the TOW design for biofilm study, we use TOW optical tweezers to disassemble clusters of *S. meliloti* cells and to show that the strength of cell-cell adhesion depends on the growth medium.

*S. meliloti* is a Gram-negative soil bacterium capable of establishing endosymbiosis with compatible host plants[35]. It serves as an advantageous model for investigating microbe–host interactions and shares with related bacteria, including those that act as pathogens, many critical factors that regulate cellular differentiation and organelle development[36,37]. Recent analysis revealed that a common laboratory strain of *S. meliloti*, Rm1021, possesses a nonsense mutation in the *podJ* gene, which encodes a conserved polarity factor that influences various cell envelope-associated functions[38]. Correcting the mutation and restoring the gene to that seen in environmental isolates of *S. meliloti* resulted in a strain that develops robust biofilms in select liquid medium. This *podJ*$^+$ strain forms cellular clusters that resemble those of related alphaproteobacteria, suggesting that the strain synthesizes an adhesive organelle at one pole of the cell, similar to the holdfast of *Caulobacter crescentus* and unipolar polysaccharide of *Agrobacterium tumefaciens*[39-41]. The amount of biofilm formed by the *podJ*$^+$ strain depends on the growth medium and correlates with the extent of cellular aggregation observed (Fig. 2). Specifically, the strain forms large cellular clusters and heavy biofilm in PYE medium, while the clusters tend to be smaller and the biofilm lighter in TY medium. There is no or very weak biofilm formation and aggregation in LB medium. Thus, the degree of biofilm formation appears to reflect the strength of intercellular attachment.

With the TOW optical tweezers, we can administer an adhesive strength assay by directly trapping and stretching a cellular cluster to estimate the underlying force. Typical experimental results are presented in Fig. 3. As shown in the left panels of Fig. 3 (a-f), cells attached to one another in the TY medium could be trapped by the TOW tweezers, stretched gradually from two ends, and, most importantly, broken apart eventually. Note that the whole disassembly process does not need any tethering or mechanical movement, nor does it require recalibration of trapping power with beam positions. The lateral pulling force can be "tuned" merely by varying the trapping beam power at the focal point and/or SLM-controlled angle and the spacing of the TOW beams. (For the results shown in Fig. 3, the length of the two stripe-like beams was about 1.5 μm each, while the spacing between the two beams was about 5 μm). The beam power used to break apart the *S. meliloti* clusters in TY medium was only about 20 mW. This simply cannot be achieved with conventional dual-trap tweezers[42] or an optical stretcher created with two counter-propagating beams[43]. By reconfiguring the TOW beams, even an asymmetrically-shaped cellular cluster in TY medium was disassembled (Figs. 3d-3f). In contrast, *S. meliloti* clusters formed in PYE medium remained intact even when the power of the TOW beam was increased to more than 5 times higher (Figs. 3g-3h), indicating stronger adhesion among the cells. From our estimate, the pulling force required to break up a *S. meliloti* cluster in TY medium should be at least 5 pN from each side of the TOW tweezers. A much stronger force would be needed in PYE medium for similar disassembly. These results illustrate that the TOW tweezers allow us to characterize quantitatively the effects of growth media on cell-cell adhesion, thus facilitating the elucidation of environmental factors that affect bacterial aggregation and biofilm formation.

We now discuss the forces involved in TOW tweezers. Direct measurement of the forces from the TOW tweezers acting on the bacteria is challenging since the forces involved in separating and trapping bacterial cells are of vastly varying magnitudes. In order to have a better understanding of the lateral forces in the TOW optical trap, we utilized a few different measurement techniques. Theoretically, in contrast to the familiar case of a spherical particle in the single-beam trap, precise calculation of stiffness and strength of compound traps such as the TOW tweezers is very complicated. In fact, only recently have theoretical models been put forth for the study of trapping forces with asymmetrical-shaped particles, and for shape-induced force fields in general[44]. The non-conservative nature of the optical force fields certainly manifests itself directly in the stiffness of trapped aspherical objects, such as a rod-shaped bacterium.

First, to substantiate that an outward pulling or splitting force indeed exists in the TOW trap, we used the method of particle image velocimetry to estimate the magnitude and direction of the flow of suspended polystyrene beads driven by the TOW tweezers. To illustrate the concept, the two beams (diverging in the x-direction) constituting the tweezers have a large separation of about 5 μm at the trapping plane, and the beads have an average size of about 500 nm. As seen in Figs. 4a-4b, the TOW tweezers behaves as two micro-pumps for a thin sample of aqueous suspension of the beads: the beads flow away from the central region along two opposite directions due to the scattering force exerted by the diverging beams (see Supplementary Movie 3). The time-averaged velocity of the particle flow is replotted in Fig. 4b, where arrowed lines mark the particle flow velocity distribution. Clearly, this diagram of hydrodynamic particle flow illustrates a transverse momentum leading to pulling in opposite directions, giving rise to the TOW action mediated by our judiciously shaped optical beam shown in Fig. 1d. Moreover, since the hydrodynamic driving of particles infers the direct relation to the force, it also provides information about the force distribution (magnitude and direction) from the TOW tweezers. As seen in Figs. 4a-4b, the pulling forces drive the particles towards the two sides rather than the central region between the two traps.

Second, to obtain an estimate of the magnitude of the trapping force from the TOW tweezers, we analyzed the time-dependent positions of a single *S. meliloti* cell trapped by only one side of the TOW tweezers using the established method of "optical potential analysis" [45]. This position distribution of the trapped cell is obtained by extracting data from video microscopy using a particle tracking software[46]. From the occupancy probability and by employing the Boltzmann statistics theorem, the potential energy, and thus, the force distribution in space can be deduced[47]. Figure 4c shows the position distribution of the cell when it is trapped by only one arm of the TOW tweezers. As seen in Fig. 4c, force distribution in the transverse y-direction (i.e., perpendicular to the pulling direction) is not notably different from that resulting from a standard Gaussian trap; in contrast, in the x-direction, the distribution becomes highly asymmetric. In other words, even when just one arm of the TOW tweezers is present, the bacterium experiences a net force in a preferred direction (i.e., along the pulling direction). Although the bacterium is not necessarily stably trapped at the point where the peak force is applied, we can still obtain a value for the peak force by fitting experimental data to the theoretical model. The results from a theoretical estimate of the force are also plotted in Fig. 4c for comparison, which gives a force of about 0.35 pN per 0.1 μm. With this method, the peak force from one arm of the tweezers is estimated to be at least 5 pN for a rod-shaped cell approximately 1.5 μm in length when displaced to the center of the TOW trap. Of course, when both traps are present under the TOW action as shown in Fig. 3a-3c for breaking up the *S. meliloti* cluster, the actual pulling forces from both sides could be much larger than this estimated value.

Third, we provide a theoretical analysis of the forces mediated by the optical landscape of the TOW tweezers. Since the beam shaped by the SLM is strongly focused by an objective lens to achieve a high field gradient, it is necessary to use a rigorous vectorial electromagnetic (EM) treatment to facilitate the modeling. As such, the Debye-Wolf integral is used to construct the field in the focal plane[47,48]. Following a similar method used previously for SLM-assisted beam shaping[47,49], the radiation in the focal plane is calculated as the integral of spherical vectors waves emanating from the objective lens. To simplify the calculation, we treat the trapped bacterium approximately as a spherical particle with a size of 1 μm and a refractive index of 1.38. The forces acting on the "particle" by the TOW beam are calculated via the T-matrix method[50,51], derived from the generalized Lorenz-Mie theory: at each point of the trapping plane, the incident EM field around the particle is calculated based on the actual holograms used experimentally on the SLM and the optical parameters of the system, and the scattered field is then calculated via the T-matrix. By comparing the incident and scattered fields, the forces on the particle can then be found by integrating fields around the particle. A typical calculated force distribution around such a particle at the trapping plane is plotted in Fig. 4d, showing a clear pulling effect on the trapped particle: The optical force is particularly strong in the middle of the TOW duel where particles will be strongly pushed away from the center of the beam.

In order to categorize the forces at work in the TOW duel, the Q value (trapping efficiency) for the optical trap is calculated. This corresponds to the amount of power required to provide a particular trapping force for a given particle. Our analyses show that, under the experimental conditions, the peak Q-value of the TOW beam is 1.5 times larger than that of a similarly positioned Gaussian trap. In other words, in TOW tweezers, a trapped particle will experience a pulling force 1.5 times greater that in Gaussian beam-based tweezers. In addition, the TOW beam has a stronger force differential between the positive and negative sides of the trap when compared with a Gaussian equivalent, leading to enhanced lateral pulling forces. Nevertheless, due to the elongated shaping of the TOW beam, the intensity threshold for bacterial photodamage could be higher as compared with that for Gaussian traps. The range of forces that the trap can exert is bounded on the lower end by the thermal force (in the low femto-newton regime, depending on temperature), and on the upper end by undesirable photodamage of bacterial cells (in the high pico-newton regime, which typically occurs at a relatively high power level, depending on the species and the trapping wavelength).

Finally, to highlight the difference between conventional dual-beam tweezers and our TOW tweezers, two approaches are used. One is to compare the position distribution of a *S. meliloti* cell trapped by only one side of the dual-traps, and the other is to compare the stability of a micro-rod trapped by both sides with two different tweezers systems. In both cases, a strongly oscillating environment[52] is provided to test the stability of the traps and to simulate ambient perturbation for motile bacteria. Although both types of tweezers can trap and hold a rod-shaped object or a bacterial cluster in the observing plane, the TOW system has obvious advantages. First, unlike in a single trap based on a symmetric Gaussian beam, a trapped object in one side of the TOW beam has a preferred direction of displacement due to its asymmetrically shaped intensity profile (see Figs. 5a-5d). As shown in Figs. 5b and 5d, under a periodic perturbation (e.g., the sample is oscillated sinusoidally), a trapped *S. meliloti* cell moves around its central equilibrium position evenly in the Gaussian trap, but it shows up more in a preferred direction in the TOW trap, indicating the pulling effect from the latter configuration. Second, to stretch the trapped object, at least one of the Gaussian-beam-based dual traps has to be moved laterally, while in the TOW tweezers, one does not need to translate the objective lens, and the object still experiences a constant stretching force due to the asymmetric intensity gradient. Third, an object trapped by TOW tweezers exhibits increased stability and resistance to ambient perturbation when compared against conventional dual-beam tweezers.

To better illustrate the advantage in trap stability, a rigid silica micro-rod is used as a test object instead of a bacterial cell to prevent any possible damage- or deformation-induced effects. Experimental results are presented in Fig. 5e and 5f. In Fig. 5e, the position distributions of the trapped rod in both the dual-beam and the TOW tweezers are plotted, where the occupancy probability is obtained by taking 10,000 snapshots from a recorded video of the micro-rod in the trap. Clearly, in the TOW tweezers, the micro-rod is better confined in the y-direction (i.e., the direction perpendicular to the stretching direction) than in the x-direction, whereas in the dual traps based on the symmetric Gaussian beams there is no such difference. Thus, when applied to a bacterial cell or a cellular cluster, the TOW tweezers give rise to a stable trapping in the y-direction along with the flexibility to move around its equilibrium position in the x-direction, which offers an advantage for stretching. In Fig. 5f, the sample experiences sinusoidal oscillation in three dimensions, as driven by a PZT (piezoelectric transducer)-actuated vibration control, and the cut-off amplitude and frequency of oscillation at which the silica rod can no longer stay in the trap are plotted for comparison. From these results, we can see that the silica rod, while it is being "stretched" in the transverse x-direction, is much more stably trapped in the y- and z- directions in the TOW tweezers than it is in the dual Gaussian beam tweezers. Clearly, both approaches coherently show that a rod-shaped object exhibits better stability when being trapped and stretched by the TOW tweezers, as compared with the conventional dual-beam tweezers.

## 4. Conclusions

In summary, we have demonstrated that judiciously shaped light beams can stretch and even break apart bacterial clusters, leading to a simple assay for cellular adhesion. In particular, we have shown that our specially designed TOW optical tweezers can be used as an effective tool for evaluating *S. meliloti* cell adhesion under different growth conditions. We have estimated the optical forces needed to disassemble *S. meliloti* flocs and determined that the trapping stability of TOW tweezers exceeds that of conventional dual-beam tweezers. This work represents another successful example of using static optical forces and novel beam shaping to perform diagnostic mechanical tests at the cellular level, and the technique can be readily adopted for studying the mechanical properties and dynamics of various living cells[53-55]. Since cellular adhesion plays a crucial role in biofilm development, our technique suggests exciting possibilities of developing new optical tools for investigating biofilm formation and related biomedical applications. Finally, this technique might be employed in single-molecule force microscopy, for example, to stretch DNA molecules without the need for positional calibration of paired traps.

**Acknowledgements**

This work was supported by NIH and NSF. We thank R. Gautam and D. Gallardo for assistance. All authors contributed significantly to this work.

**Figure caption list:**

Fig. 1: (Left panels) Different designs of optical tweezers. (a) Single-beam optical tweezers tend to align a rod-shaped bacterium along the beam axis. (b) Dual-beam optical tweezers trap a bacterium at each end to orient the cell into the observing plane. (c) Tug-of-war (TOW) optical tweezers trap a bacterium at each end but also exert lateral forces in opposite directions. (Right panel) Composite image illustrating the process of how a bacterial cluster is trapped, stretched, and separated by the TOW tweezers. The diagram in (d) comprises the vector field of the intensity gradient of the trapping beam (white arrows), a volumetric rendering of the beam from experimental data near the focus of an objective lens (green shading), and a schematic representation of a pair of attached bacterial cells being trapped and pulled apart. The inserts in (d) show snapshots of a dividing *B. thuringiensis* cell that was aligned gradually onto the observing plane and stretched by the TOW tweezers (see Supplementary Movie 1).

Fig. 2: *S. meliloti podJ$^+$* cells display distinct assemblages, as well as different levels of flocculation and biofilm formation, in three growth media (LB, TY and PYE). Top panels (a) show phase contrast images of the *S. meliloti* cells and aggregates, while bottom panels (b) show photographs of corresponding bacterial biofilm formed in 16-mm-diameter glass tubes.

Fig. 3: (a-c) Snapshots of a trapped *S. meliloti* aggregate consisting of four attached, self-aligned cells in TY medium, being gradually stretched from two ends and eventually broken up into two pieces by the TOW optical tweezers. (d-f) Similarly dynamic process for the disassembly of an asymmetrically-shaped *S. meliloti* cluster grown in TY medium (see Supplementary Movie 2). (g, h) Strong binding of several *S. meliloti* cells in PYE medium into a cluster (g), which remains intact under the action of the TOW tweezers (h).

Fig. 4: (a) A snapshot of suspended polystyrene beads (500 nm in diameter) driven laterally to two opposite sides by the TOW tweezers. The locations of the two main intensity spots from the "diverging" TOW beams are marked by two dashed circles, from where the beads are pushed away (see Supplementary Movie 3). (b) Illustration of time averaged direction and magnitude of the flowing beads when the two beams constituting the TOW tweezers are 5 μm apart at the focal plane. The color represents the magnitude of the normalized average particle velocity, and arrows indicate the direction of particle flow. (c) Trapping force resulting from only one arm of the TOW tweezers while the other is absent (i.e., that part of the beam is blocked). Measured results are plotted in solid curves with shaded area representing the error. The theoretically calculated force in x-direction is shown on the same graph (dashed curve) for comparison. (d) The force profile of a 1-μm "bacterium-like" particle with a refractive index 1.38 calculated via the generalized Lorenz-Mie theory. Red and blue colors represent the magnitude of the force in positive and negative directions, respectively. The overlay shows the forces (in normalized units) along the dotted line. The hair-like wisps represent paths taken by simulated particles in the absence of Brownian motion. The dotted box indicates the region similar to that shown in (c).

Fig. 5: Comparison between conventional dual-beam tweezers and TOW tweezers. (a) Intensity profile of the Gaussian dual-beam trap. (b) Position distribution of a *S. meliloti* cell trapped by only one side of the dual-trap in an oscillating environment (i.e., the sample is sinusoidally dragged at a frequency of 20Hz and an amplitude of 2μm), overlaid with the intensity profile of the trap (in pink). (c, d) Corresponding results for the TOW tweezers. (e) Occupation probability of a trapped silica rod around its equilibrium position in the two different traps. With the TOW tweezers, the micro-rod is more stably trapped in the y-direction yet still has the flexibility to move around its equilibrium position in the x-direction (desirable for stretching). Inserts show the location of each beam on the silica rod. (f) Plot of maximum oscillating amplitude and frequency at which the silica rod can stay in the three-dimensional trap, under the action of either dual-beam or TOW tweezers. Enhanced stability in the TOW tweezers is evident.

## Additional information

All authors contributed to this work. The authors declare no competing financial interests. Supplementary information is available in the online version of the paper. Correspondence and requests for materials should be addressed to Z.C.

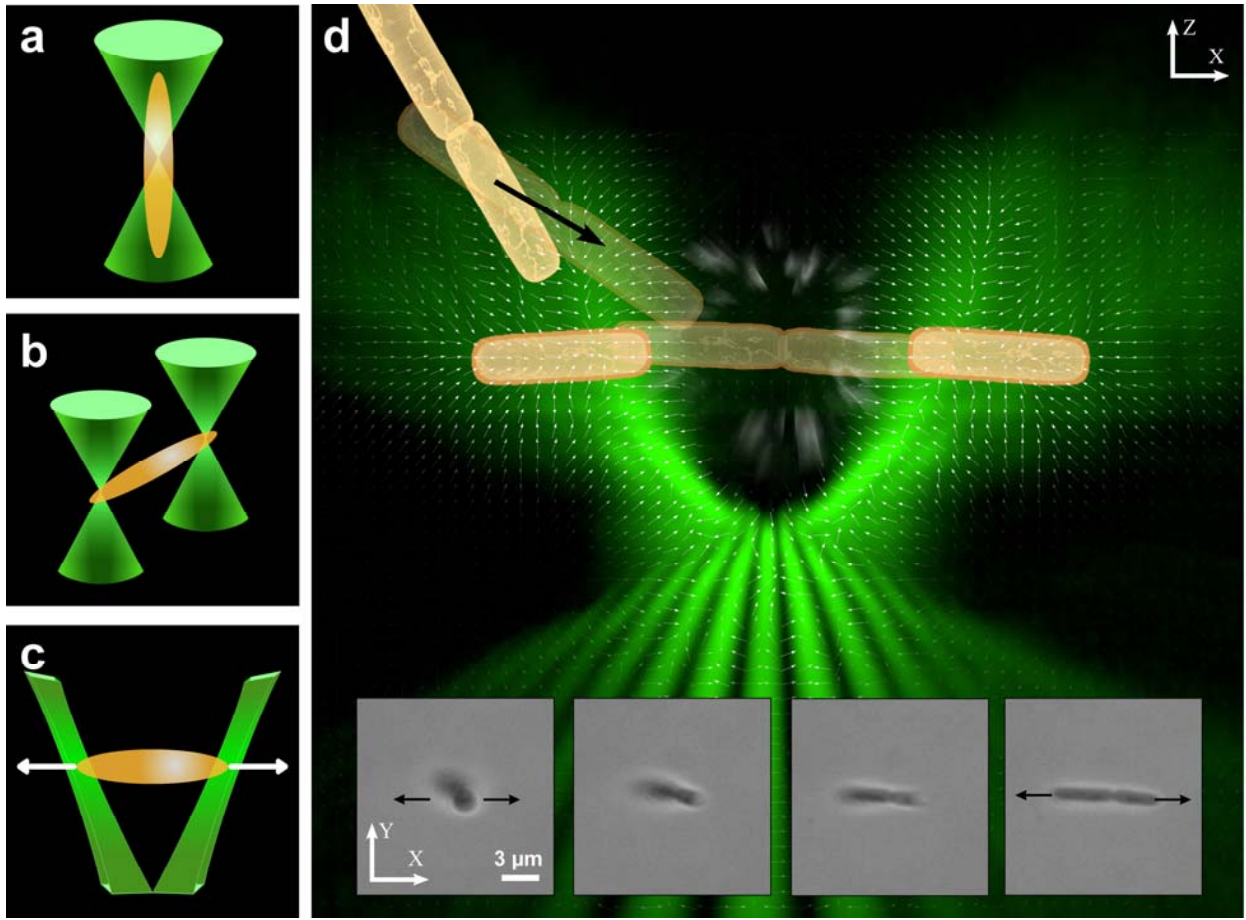

Fig. 1: (Left panels) Different designs of optical tweezers. (a) Single-beam optical tweezers tend to align a rod-shaped bacterium along the beam axis. (b) Dual-beam optical tweezers trap a bacterium at each end to orient the cell into the observing plane. (c) Tug-of-war (TOW) optical tweezers trap a bacterium at each end but also exert lateral forces in opposite directions. (Right panel) Composite image illustrating the process of how a bacterial cluster is trapped, stretched, and separated by the TOW tweezers. The diagram in (d) comprises the vector field of the intensity gradient of the trapping beam (white arrows), a volumetric rendering of the beam from experimental data near the focus of an objective lens (green shading), and a schematic representation of a pair of attached bacterial cells being trapped and pulled apart. The inserts in (d) show snapshots of a dividing *B. thuringiensis* cell that was aligned gradually onto the observing plane and stretched by the TOW tweezers (see Supplementary Movie 1).

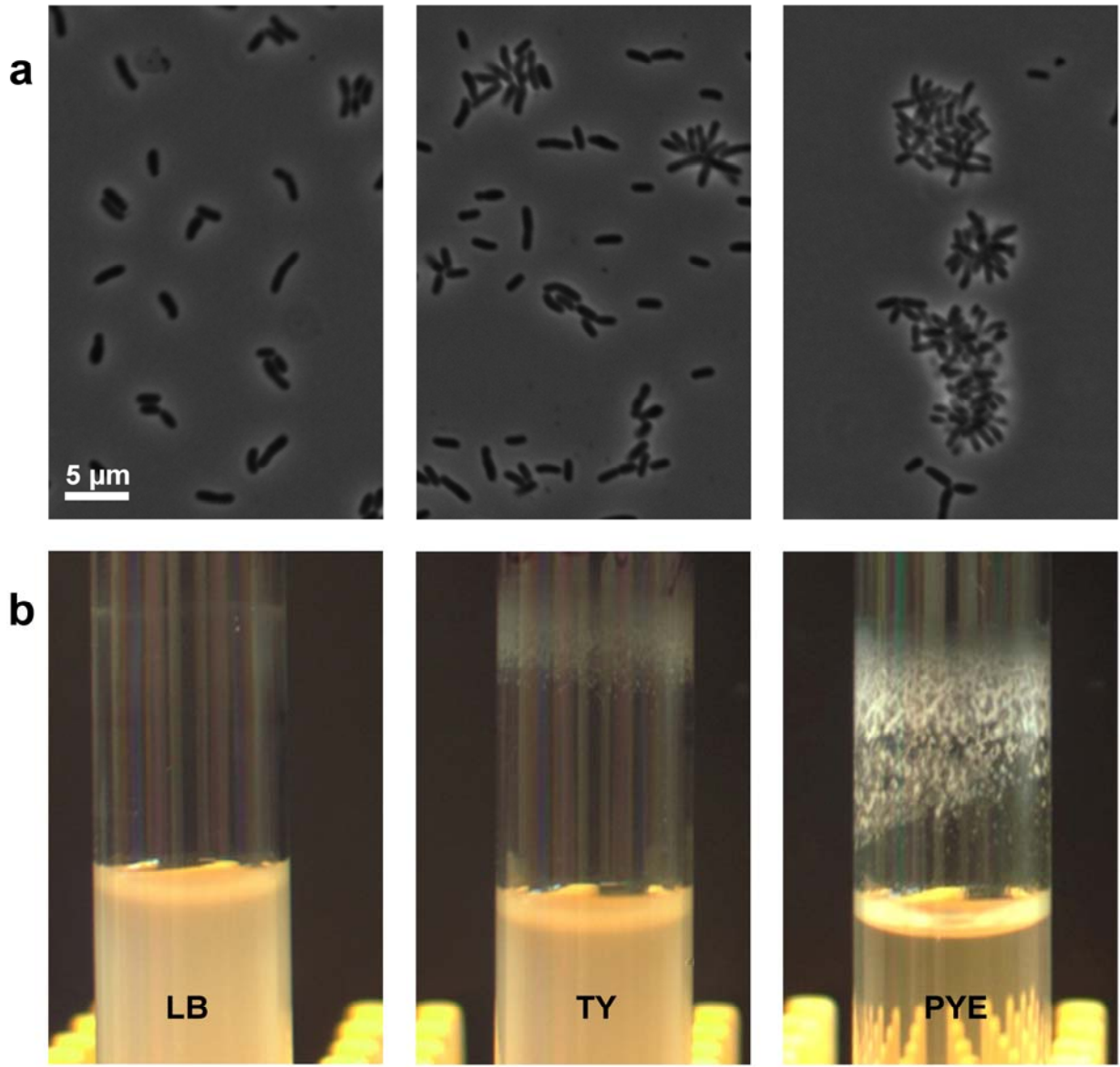

Fig. 2: *S. meliloti podJ*[+] cells display distinct assemblages, as well as different levels of flocculation and biofilm formation, in three growth media (LB, TY and PYE). Top panels (a) show phase contrast images of the *S. meliloti* cells and aggregates, while bottom panels (b) show photographs of corresponding bacterial biofilm formed in 16-mm-diameter glass tubes.

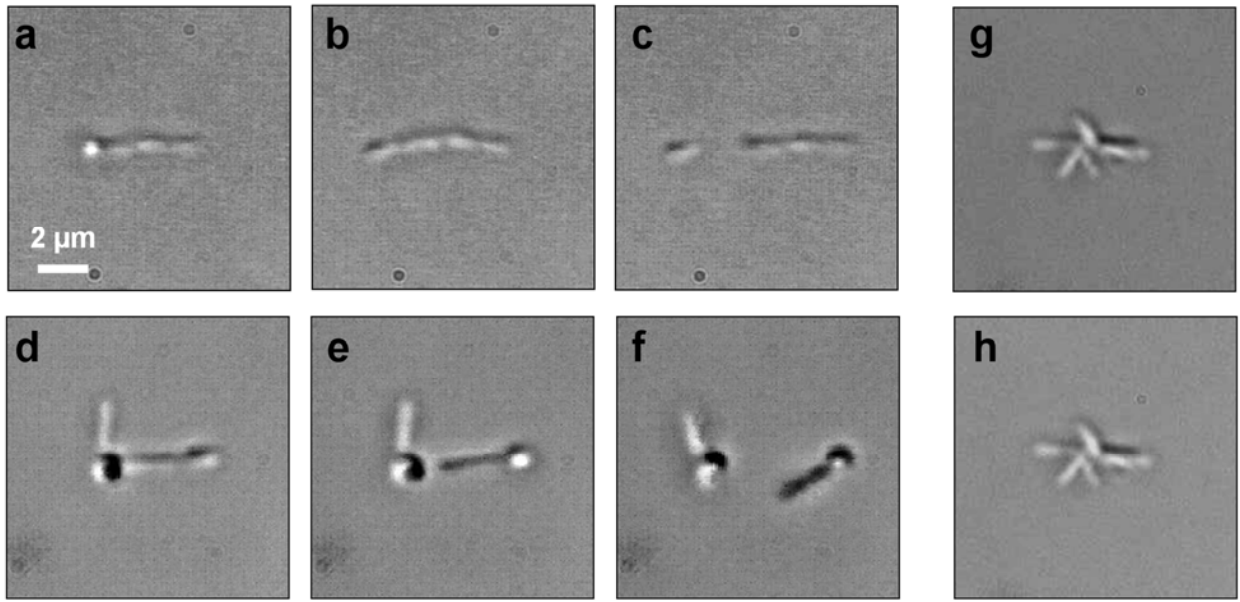

Fig. 3: (a-c) Snapshots of a trapped *S. meliloti* aggregate consisting of four attached, self-aligned cells in TY medium, being gradually stretched from two ends and eventually broken up into two pieces by the TOW optical tweezers. (d-f) Similarly dynamic process for the disassembly of an asymmetrically-shaped *S. meliloti* cluster grown in TY medium (see Supplementary Movie 2). (g, h) Strong binding of several *S. meliloti* cells in PYE medium into a cluster (g), which remains intact under the action of the TOW tweezers (h).

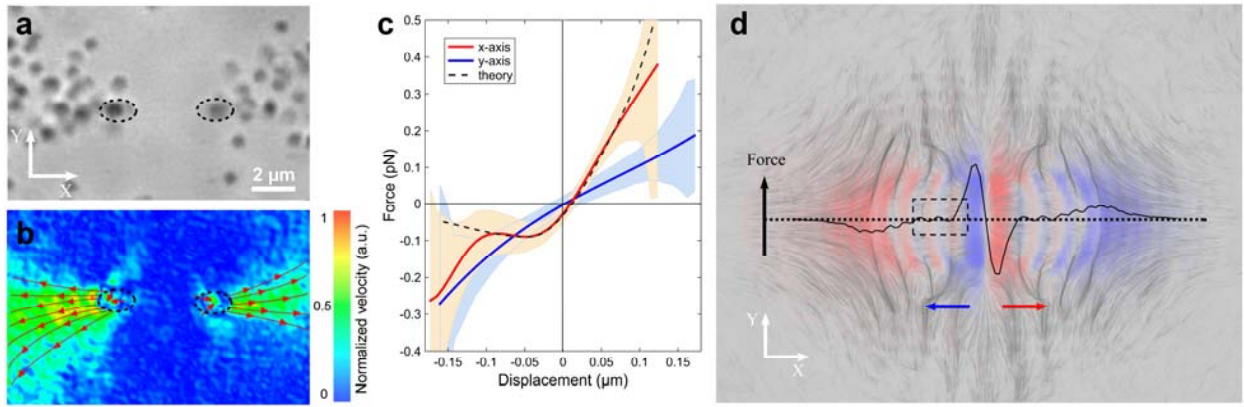

Fig. 4: (a) A snapshot of suspended polystyrene beads (500 nm in diameter) driven laterally to two opposite sides by the TOW tweezers. The locations of the two main intensity spots from the "diverging" TOW beams are marked by two dashed circles, from where the beads are pushed away (see Supplementary Movie 3). (b) Illustration of time averaged direction and magnitude of the flowing beads when the two beams constituting the TOW tweezers are 5 μm apart at the focal plane. The color represents the magnitude of the normalized average particle velocity, and arrows indicate the direction of particle flow. (c) Trapping force resulting from only one arm of the TOW tweezers while the other is absent (i.e., that part of the beam is blocked). Measured results are plotted in solid curves with shaded area representing the error. The theoretically calculated force in x-direction is shown on the same graph (dashed curve) for comparison. (d) The force profile of a 1-μm "bacterium-like" particle with a refractive index 1.38 calculated via the generalized Lorenz-Mie theory. Red and blue colors represent the magnitude of the force in positive and negative directions, respectively. The overlay shows the forces (in normalized units) along the dotted line. The hair-like wisps represent paths taken by simulated particles in the absence of Brownian motion. The dotted box indicates the region similar to that shown in (c).

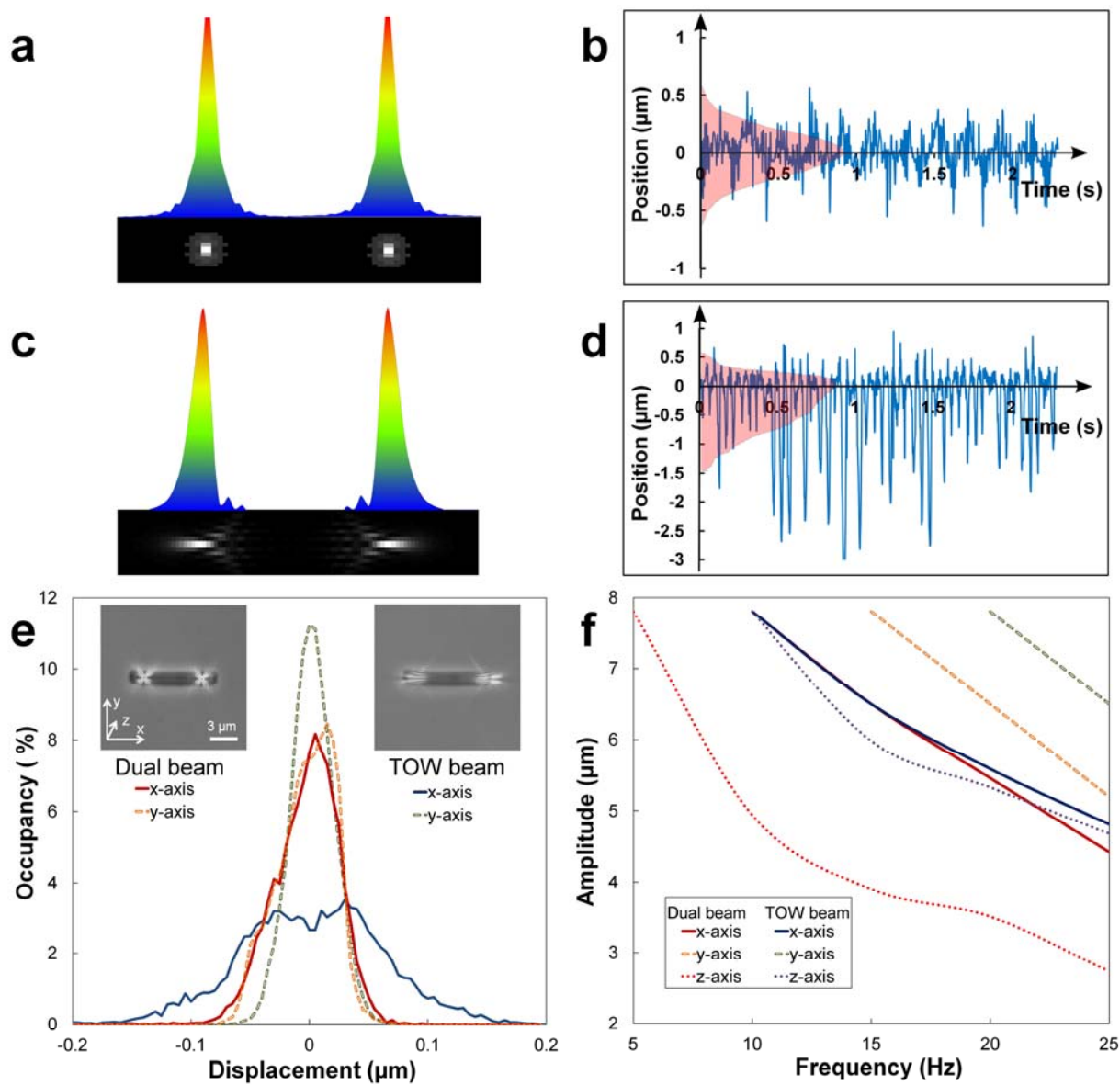

Fig. 5: Comparison between conventional dual-beam tweezers and TOW tweezers. (a) Intensity profile of the Gaussian dual-beam trap. (b) Position distribution of a *S. meliloti* cell trapped by only one side of the dual-trap in an oscillating environment (i.e., the sample is sinusoidally dragged at a frequency of 20Hz and an amplitude of 2μm), overlaid with the intensity profile of the trap (in pink). (c, d) Corresponding results for the TOW tweezers. (e) Occupation probability of a trapped silica rod around its equilibrium position in the two different traps. With the TOW tweezers, the micro-rod is more stably trapped in the y-direction yet still has the flexibility to move around its equilibrium position in the x-direction (desirable for stretching). Inserts show the location of each beam on the silica rod. (f) Plot of maximum oscillating amplitude and frequency at which the silica rod can stay in the three-dimensional trap, under the action of either dual-beam or TOW tweezers. Enhanced stability in the TOW tweezers is evident.